\documentclass{webofc}
\usepackage[varg]{txfonts}   
\pdfoutput=1 

\usepackage{pifont}
\usepackage{enumitem}
\usepackage{longtable}

\RequirePackage{lineno}
\usepackage{rotating}
\usepackage{graphicx}
\usepackage{graphics}

\usepackage{subfigure}
\usepackage{wrapfig}
\usepackage{dcolumn}
\usepackage{bm}
\usepackage{longtable}
\usepackage[percent]{overpic}
\usepackage{color}

\usepackage{float}
\usepackage{psfrag}

\def\vf{\varphi}

\def\pt{p_{\rm T}}
\def\sNN{\mbox{$\sqrt{s_{_{\rm NN}}}$}}   
\def\av#1{\langle #1 \rangle}

\woctitle{?????????}
\begin{document}
\title{Azimuthal flow in hadron collisions from quark-gluon string repulsion}
%
%

\author{\firstname{Igor  } \lastname{Altsybeev }\inst{1}\fnsep\thanks{\email{Igor.Altsybeev@cern.ch}} and 
{\firstname{Grigory } \lastname{Feofilov }\inst{1}\fnsep\thanks{\email{feofilov@hiex.phys.spbu.ru }}}
}

\institute{Saint-Petersburg State University}

\abstract{%
Color flux tubes (quark-gluon strings), formed at early stages of hadron-hadron collisions, may overlap in case of sufficiently high densities
and interact, producing long-range azimuthal correlations.  
In the hypothesis of repulsive interaction, 
each string may acquire, before the hadronization, the additional transverse boost, 
which is an efficient sum of all accounted string-string interactions. 
This 
modifies transverse momenta to the particles formed in string decay, 
leading to modification of 
event-wise 
observables, like azimuthal asymmetry of two-particle correlations, over a wide range of rapidity. 

In this article we discuss results of 
Monte Carlo model with string repulsion, where 
efficient string-string interaction radius is introduced.
We show that  the effect of string repulsion 
can be the main dynamic origin of the elliptic flow and of the higher harmonics, 
which are reflected in the complicated structures observed in two-particle long-range correlation topology in
nucleus-nucleus collisions at RHIC and at LHC.

}
\maketitle

\section{Introduction}
\label{sec-1}

The very first proposals for studies of azimuthal and long-range correlations were formulated in \cite{Abramovsky-1980}. It was emphasized  that the data on the azimuthal asymmetry in soft multiparticle production may contain very non-trivial information. 
In  \cite{ellliptic-1992} it was shown that anisotropies in transverse-momentum distributions provide an unambiguous signature of transverse collective flow in ultrarelativistic nucleus-nucleus collisions.

The azimuthal anisotropy in two-particle correlation functions observed at RHIC and LHC is 
characterized by the 
Fourier flow coefficients \cite{Voloshin-1996, Voloshin-1998}:

 \begin{equation}
\label{vn}
v_{n}=\av{\cos[n(\varphi-\Psi_{n})]}.
\end{equation}
Here $\varphi$ is the azimuthal angle of the particle, $\Psi_{n}$ is the
angle of the initial state spatial plane of symmetry, and $n$ is
the order of the harmonic. The planes of symmetry
$\Psi_{n}$ are not known experimentally and the anisotropic flow
coefficients are estimated from measured correlations between
the observed particles. The second Fourier coefficient
$v_2$ is called elliptic flow and has been studied in details
in recent years in number of experiments.
The  first results on the elliptic flow of charged particles at midrapidity 
in Au+Au collisions at
130~GeV were reported  by the STAR Collaboration  at RHIC \cite{STAR-2001}. 
It was assumed that 
the observed azimuthal anisotropy appears at very early time and 
is caused by the pressure gradient and  anisotropic expansion of the  matter formed in the collisions 
\cite{ellliptic-1992,Voloshin-1996,Voloshin-1998,Sorge-1997}. 
The anisotropy effects 
were confirmed later 
in Pb+Pb collisions at the LHC.
For example, large values of elliptic flow at the LHC were 
observed by the ALICE Collaboration \cite{ALICE-elliptic}.
In addition to the elliptic flow,  direct ($v_1$), triangular ($v_3$) and  higher flow coefficients  are also obtained \cite{Voloshin-1998}.  
It was established that the initial state spatial anisotropy is converted, 
due to some 
collective behaviour,  to the momentum anisotropy. 

In the experimental studies of flow,  two main methods are usually applied. 
In the first method, flow coefficients are extracted
from a two-particle angular correlations function, usually studied 
in terms of difference between two particles in azimuthal angle $\Delta\vf$
and pseudorapidity $\Delta\eta$.
Correlations can be studied either between 'trigger' and associated particles
(which can be taken from different $\pt$ ranges and by some PID selection criteria)
or just by taking all pairs of particles in an event
(untriggered correlations).
Another method is based on multi-particle cumulants using moments of the flow vectors \cite{cumulants}, 
which allows to eliminate a contribution from correlations that are not related to the initial geometry (so-called non-flow).

A new set of striking phenomena was obtained in the two-particle correlation studies:  the so-called ''ridge''  was first observed 
in Au+Au collisions at $\sqrt{s_{NN}}=200$ GeV at RHIC \cite{STAR-2004,STAR-2005,STAR-2006}.
Another puzzling experimental effect in the two-particle angular correlations 
was obtained later in the first measurement of the triangular $v_3$ and higher order harmonics of charged particle flow in Pb-Pb collisions at $\sqrt{s}=2.76$ TeV  \cite{ALICE-H, ATLAS-H}:
a double-peak structure in the away-side azimuthal region was discovered 
in the most central events. 
It was shown in \cite{ALICE-H} that this structure can be naturally described
 by the 
anisotropic flow Fourier coefficients, in particular, by adding the higher harmonics. 
Detailed studies of  $\pt$, centrality and rapidity dependences  of  elliptic and higher order flow harmonics 
were performed  by CMS \cite{CMS-PbPb-r},  ATLAS \cite{ATLAS-PbPb-pt-dependence} and ALICE \cite{ALICE-H,ALICE-PbPb-Pseudorapidity}  at the LHC.
Recent review on  novel phenomena in particle correlations in relativistic heavy-ion collisions can be found in \cite{review-2014}.

The  azimuthal anisotropy in two-particle correlations  in Au+Au  and Pb+Pb collisions is generally related  to  manifestations of high-order flows and is  interpreted as the result of a hydrodynamic expansion of the medium (of a nearly perfect liquid) formed in the collision. We have to mention, that contrary to this interpretation, the collective flow effects  in relativistic ion collisions could be also explained  without the  hydrodynamic assumptions in so-called particle free streaming approach \cite{free streaming}. 
The last one was motivated by the fact that the effects,   similar to the collective flow  and  associated with hydrodynamics, were  not expected for elementary pp or p+Pb  collisions. 
Remarkable experimental observations were done by  CMS
in  very high multiplicity proton-proton collisions at the LHC:
a positive long-range correlation was reported between two particles produced at similar azimuthal angles, spanning a large range in rapidity   \cite{CMS-pp-2010}. 
Recent CMS results on multi-particle correlations in pp collisions 
show that this structure has collective nature \cite{CMS-pp-2016}.
The strong effects were also found in p+Pb collisions \cite{CMS-pPb}.
Also, unexpectedly for light colliding systems, the existence of so-called double-ridge structures was  found 
 in d+Au collisions at RHIC \cite{STAR- diH, PHENIX-dAu}  and 
in p+Pb collisions at the LHC 
\cite{ALICE-pPb, ATLAS-pPb}. 
These experimental findings produced a number of challenges to the existing theoretical approaches.

The characteristic feature of these azimuthal asymmetries  of multi-particle angular correlations is that they are extending  over a long range in rapidity. Causality then requires that these long-range correlations, if they exist,  originate from the earliest times of the collision \cite{Dumitru}. 
Therefore, we would like to mention here  just briefly  those  models  that are related intrinsically  to the initial stages of the QGP formation and to the observed long-range rapidity correlation phenomena (one may  see also the reviews \cite{Li} and  \cite{teor}).

In the  model of color glass condensate (CGC) the boost invariant Glasma flux tubes are considered \cite{Dumitru}  which  produce particles isotropically with equal probability along the length of the flux tube. Interactions among those particles produces pressure  leading  to collective radial flow.  The average transverse velocity 
is obtained here from the blast wave fits to the RHIC data obtained by the PHENIX collaboration. Thus the common transverse expansion experienced by particles in a flux tube collimates the particles forming a ridge like structure extended in rapidity.

The string-like approach considers   the quark-gluon strings, formed in the result of color exchange between partons of colliding hadrons, as  particle emitting sources. 
These sources are stretched longitudinally between the beam remnants and  at certain density in the transverse plane they may start to interact, forming clusters and producing some collective effects. One may mention here the string-fusion and percolation phenomena extensively studied in nucleus-nucleus collisions  in \cite{Perc_ridge, Paj-2011}.
In \cite{Perc_ridge} the string percolation phenomenology is  compared
 to the CGC results on effective string or glasma flux tube intrinsic correlations, including the ridge
phenomena and long-range forward-backward azimuthal correlations. Color string
percolation model and its similarities with the CGC were discussed in \cite{Paj-2011}.

In the theoretical analysis of long-range  correlations there are two major factors that should be  taken into account. In the string-like approach both are intrinsically present:  (i) formation of  the particle production sources extended in rapidity and  (ii) certain initial anisotropy of configuration of these sources (color-flux tubes or strings) in the transverse plane. 
It is the last factor that might be  responsible for the azimuthal anisotropy  effects of transport of particles through the given  non-uniform medium.
Among these approaches  are the recent studies \cite{Braun-1, Braun-2, Braun-3} of the azimuthal anisotropy of two-particle correlation functions, observed  in hadron collisions, that are based on the assumption of quark-gluon string formation at the very early stages of hadron collisions and further interaction of particles produced by these sources  with  the strong colour fields inside string clusters.   It is shown  by the direct Monte Carlo simulations \cite{Braun-3} that the elliptic flow can be successfully described in the color string picture with fusion and percolation, provided the anisotropy of particle emission from the fused string is taken into account.
Two possible sources of this anisotropy are considered, propagation of the string in the transverse plane and quenching of produced particles in the strong colour field of the string \cite{Braun-3}.

Another  motivation to understand the dynamics of QGP formation in relativistic heavy-ion collisions can be found in   the dual holographic approach  (one may see review in \cite{Arefeva}).
These issues of holography and AdS/CFT duality are  being intensively studied at present. The  system of QCD strings is considered to be attractively interacting and collapsing, and it is viewed as a QCD analog to AdS/CFT black hole formation 
\cite{Shuryak-1,Shuryak-2, Shuryak-3}.

In string model approach, following general assumptions  from the Regge phenomenology, it is considered that   quark-gluon strings (color flux tubes) are formed  between the partons of the colliding nuclei at the early stage of hadron-hadron collision. The number of these strings  depends on the number of  participating nucleons ($N_{part}$) and on the collision energy. In case of sufficiently high density these strings may overlap and interact by attracting or repulsing,

 Possible interaction of color strings (color flux tubes) in the form of repulsion or attraction and  formation of flow harmonics in hadron collisions  was first   considered more then 3 decades ago  in \cite{Abramovsky-1980}. The exact type of string-string  interaction is not known. The sign of string-string interactions depends on the direction of color fluxes \cite{Abramovsky-1980}. 
This problem of the string-string interaction has not yet been systematically addressed till recently~\cite{Shuryak-1,strings-interaction, Shuryak-2014}.
 The magnitude of this interaction in string tension units was found to be small $\sim 10^{-1}-10^{-2}$~\cite{strings-interaction}. It is the collective effect of the large number of strings that is responsible for  significant compression of the system in the transverse plane, leading in case of attractive interaction to implosion of such high-density configurations of  strings  \cite{Shuryak-2015}. 

Recently a new interest to interactions bewteen color flux tubes appeared, that was  motivated both by the experimental findings of long-rage correlation phenomena in hadron collisions and by the progress  in  fitting  the data in the framework of the  various string models. They include Pomeron  exchange approach,  Lund model and  MC event generators based on this model,  the  collectivity effects like color reconnection and string fusion and percolation phenomena, etc. 
One may conclude that the field of initial stages of hadron collisions, in particular, in string-like approach, is being extensively explored.

 In the present study we continue our investigations \cite{OK,IA,QC2014,OK-IA-GF} of  the role of string density effects 
in the formation of azimuthal asymmetry of two-parrticle correlation functions in the case of string-string repulsion.
The paper  has the following structure.  After the Introduction we describe in the Section~2  the approach and the main assumptions that are implemented in our  
MC model \cite{IA}. 
Two-particle correlation functions at different centralities and harmonic decomposition of the azimuthal correlation function, obtained in the MC model, are presented in Section 3.
Results on the $\pt$ spectra and mass-splitting of the elliptic flow coefficients are  presented in the Section 4.
Discussion and conclusions are given in the last two sections.


\section{MC model with repulsive  strings }
\label{sec-2}

In our study we consider a simplified approach to string-string interaction mechanism for the case of repulsion. 
Basing on this hypothesis,
a Monte Carlo (MC)  model \cite{IA} was developed.
In this MC model, final-state particles from A+A interactions are generated in each event in the following stages:

\noindent {\it Stage 1. Simulation of the nuclei.}\\
Initial positions of the nucleons in nuclei are generated in accordance to Woods-Saxon distribution.
Nuclei of Pb$^{208}$ are considered, the Woods-Saxon radius is 6.62 fm, diffusion  = 0.546 fm. Inside each nucleon partons are distributed in transverse ($xy$) plane with 2D-Gauss law with $\sigma_{xy} = 0.4$ fm.
Number of partons
inside each nucleon is generated by Poisson law,
where the mean number depends on a collision energy and is a model parameter. Two such a nuclei with some impact parameter
are generated in each collision.

\noindent {\it Stage 2. Simulation of strings configuration.}\\
Interaction between colliding hadrons is implemented at the partonic level:
partons from colliding nuclei can interact forming a qurk-gluon string, if the distance between partons in $xy$ plane is less then 
parton interaction distance $d=0.4$ fm (another model parameter).
Additionally, there is a 3\% probability for a string to be a result of ''hard scattering''  of partons to emulate of jet-like structures. 
Other 97\% 
of the strings are  considered to be "soft" and long in rapidity,  occupying rapidity range  $y\in(-4,4)$ .

\noindent {\it Stage 3. Repulsion of the strings.}\\
At the next step of the system evolution,  "soft" strings  interact with each other.
In the current MC model, interaction 
manifests itself as a repulsion.
The  repulsion mechanism is adopted from \cite{Abramovsky-1980}. 
In our MC model, an efficient string-string interaction radius $R_{int}$ is introduced in the transverse plane. We consider this variable $R_{int}$ as a free parameter (it is different from the string radius $r_{0}$ which is often considered to be of about $r_0=0.2$ fm).
$R_{int}$ determines "interaction zone" around each string,
these zones may overlap,  and, due to string-string mutual repulsion, in case of high string density each individual string
acquires significant transverse boost $\beta$ 
(see \cite{Abramovsky-1980} and \cite{IA} for detailed mechanism). The mean value of the string boost is adjusted to be 
 $\overline{\beta} \approx 0.65$ in central Pb-Pb collisions. 

\noindent {\it Stage 4. Event final state: hadronization of the strings.}\\
Each string breaks in several places
into quark-antiquark pairs
with the same exponentially distributed $\pt$
and in
opposite azimuthal directions.
Quarks from neighbour  string pieces then combine and form a meson (mostly pions and  $\rho$). 
{\it In the laboratory frame, all the particles originated from one string  are boosted by the factor $\overline{\beta}_{\rm string}$.}
Due to the numerous string-string interactions taken into account, the initial configuration of quark-gluon strings could be transferred into the particles azimuthal flows.

The modified version of the MC model  will be used in Section \ref{sec-3} to consider spectra and mass ordering 
for the particles with different masses.

\section{Some features of MC model with string repulsion}

\begin{figure}[h]
\centering
$\begin{array}{ccc}
\begin{overpic}[width=0.32\textwidth, clip=true, trim=0 0 70 0]
{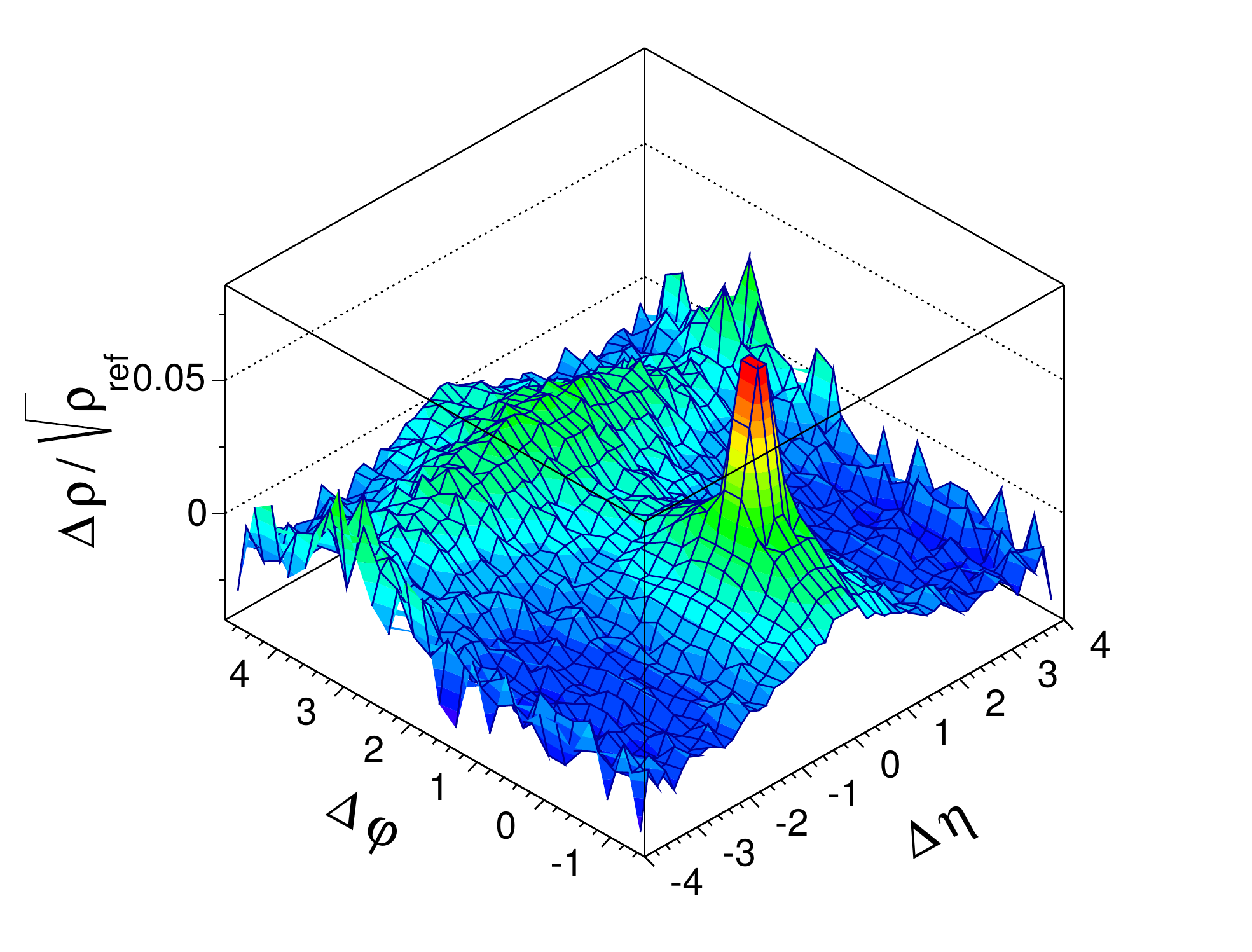}
	\put(2,86){  peripheral events}
	\put(2,78){  $\pt>0.15$ GeV/c } 
\end{overpic}
&
\begin{overpic}[width=0.32\textwidth, clip=true, trim=0 0 70 0]
{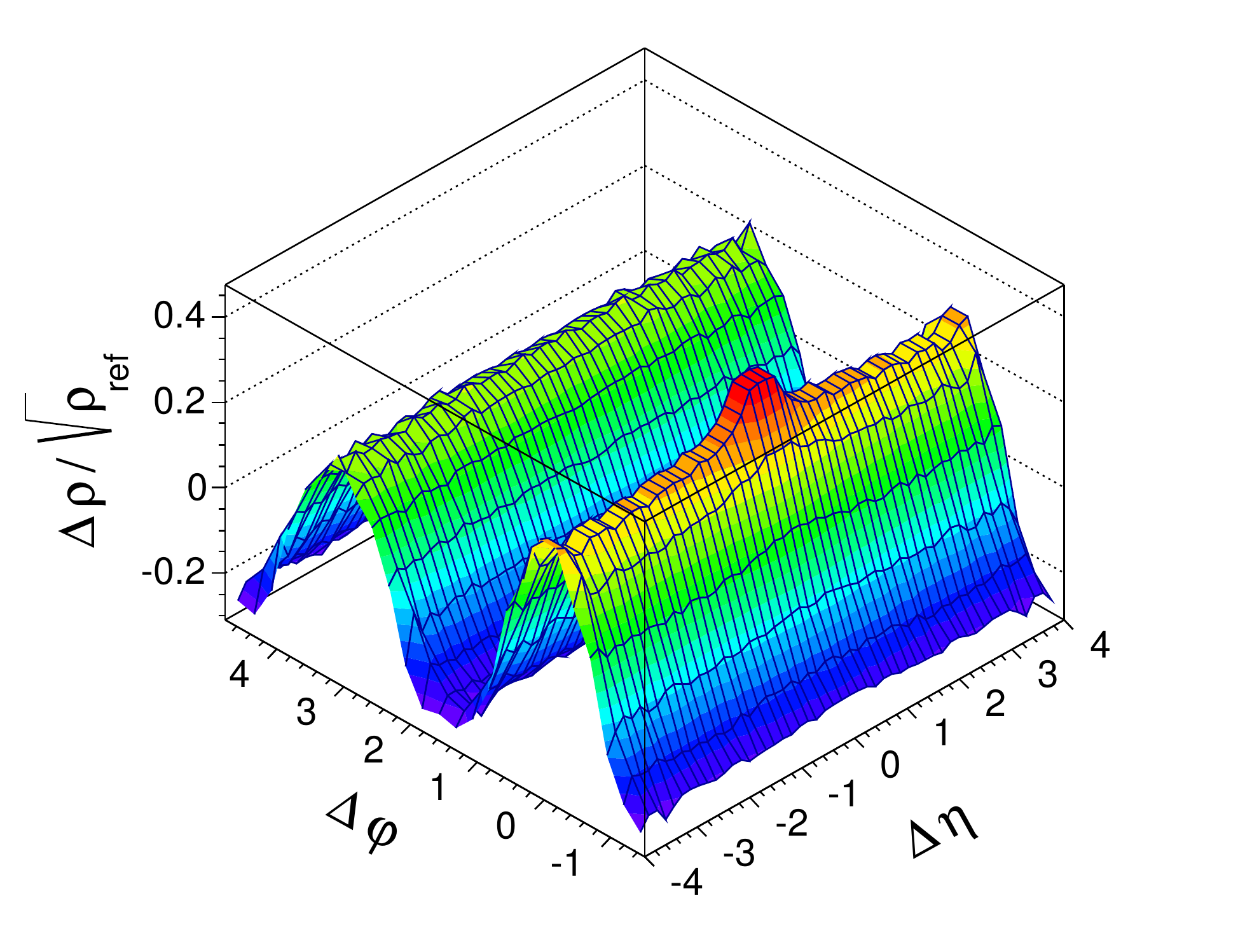}
	\put(2,86){  semicentral events}
	\put(2,78){  $\pt>0.15$ GeV/c }
\end{overpic}
&
\begin{overpic}[width=0.32\textwidth, clip=true, trim=0 0 70 0]
{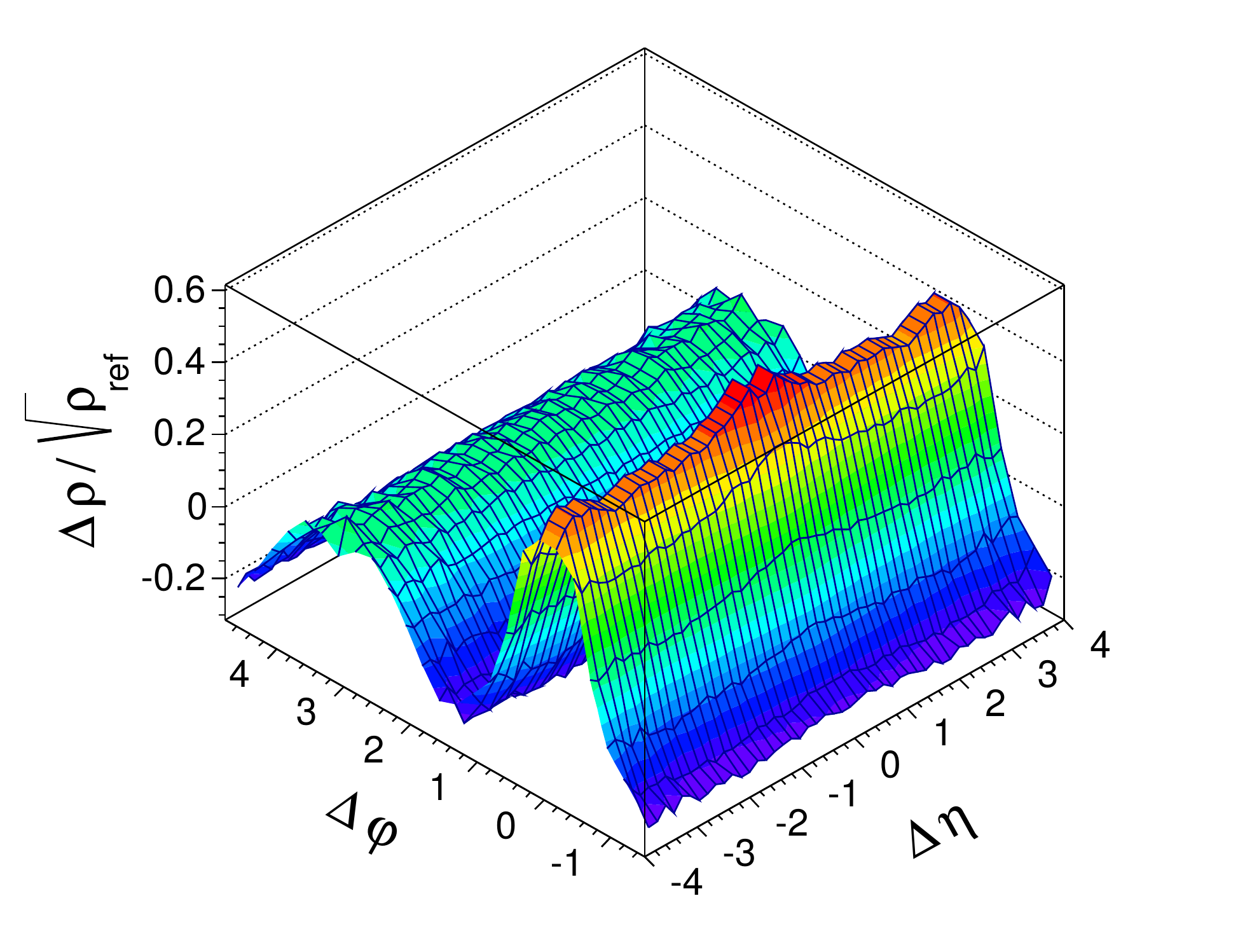}
	\put(2,86){  central events}
	\put(2,78){  $\pt>0.15$ GeV/c }
\end{overpic}
\end{array}$
\caption{
MC model~\cite{IA} results for two-particle correlation functions obtained in the
for peripheral  (left), semi-central (middle) and  central  (right) events (from \cite{QC2014}).
Particles with $\pt>0.15$ GeV/{\it c} are analyzed.
String-string  interaction radius is 2 fm.
}
\label{fig:dihadron}
\end{figure}

It was shown earlier \cite{QC2014} that the MC model with string repulsion  is capable to reproduce the characteristic features  of two-particle correlation functions obtained in  peripheral, semi-central and  central   collisions  of heavy ions 
for charged particles, 
if rather large  string-string  interaction radius $R_{int}=2$~fm is used (see  Figure\,\ref{fig:dihadron}). 

The pads in the Fig.~\ref{fig:dihadron}  illustrate the onset of collectivity with the increase of string density when passing from peripheral to central A-A collisions.
In peripheral events, a structure along $\Delta\varphi$ is visible, which is formed 
due to the $\rho^0$  decays into pions. One may see also, that the evolution of two-particle angular correlations with centrality  and the relevant transition from low-density to the high density in Au-Au collisions at $\sqrt{s_{\rm NN}}$ = 200 GeV  is correctly reproduced. (Note that STAR  variables $\Delta{\rho}/\sqrt{\rho_{\rm ref}}$ \cite{STAR- diH} are used here for two-particle correlations.)

\vspace*{0.2cm}


\begin{figure}[h]
\centering
\subfigure[a][]
{
\begin{overpic}[width=0.33\textwidth, clip=true, trim=10 0 61 0]{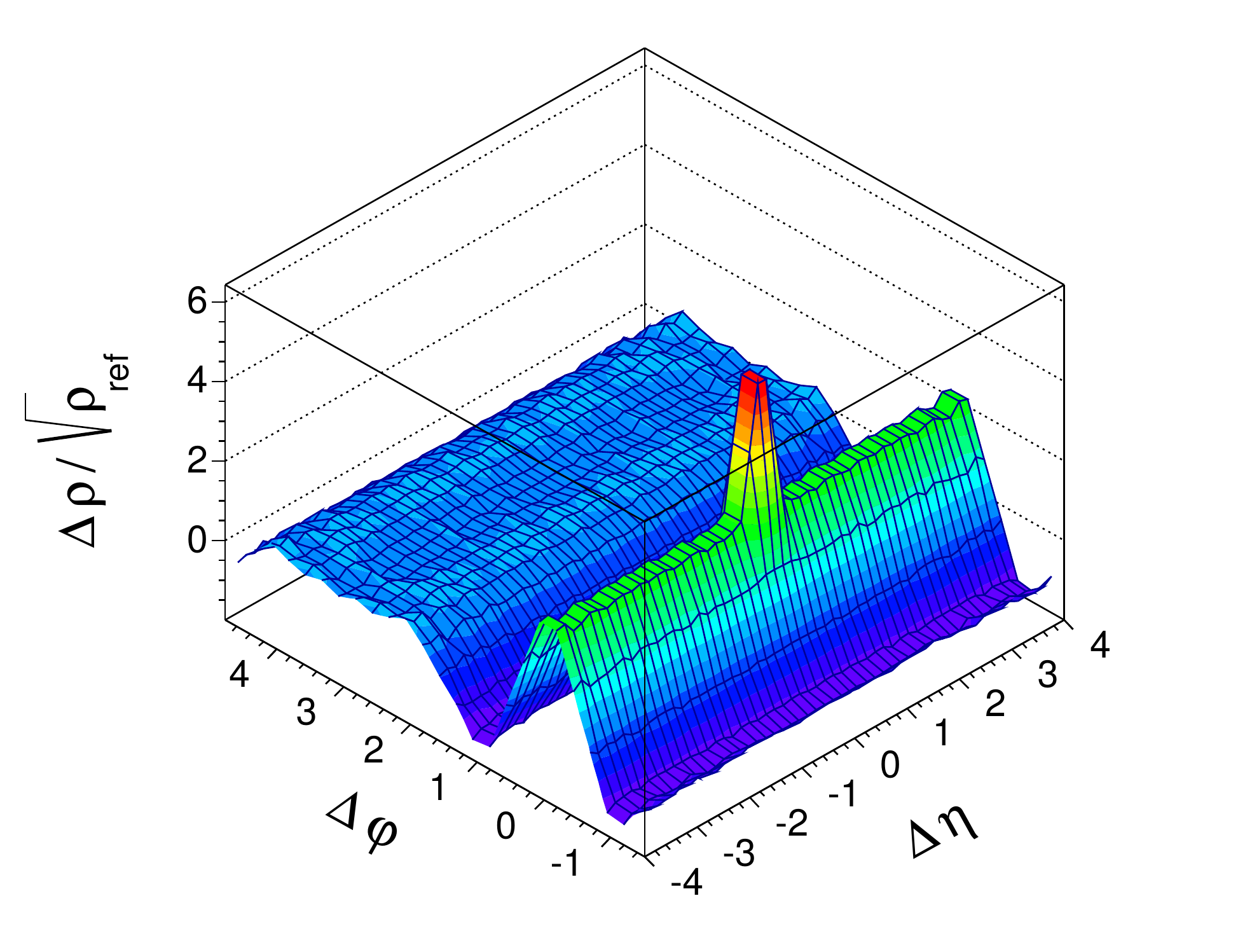}
      \put(60,82){  MC model }
	\put(0,82){ central events}
	\put(0,74){ $\pt$ 3-5 GeV/c }
\end{overpic}
}
\hspace{-0.4cm}
\subfigure[a][]
{
\begin{overpic}[width=0.33\textwidth]{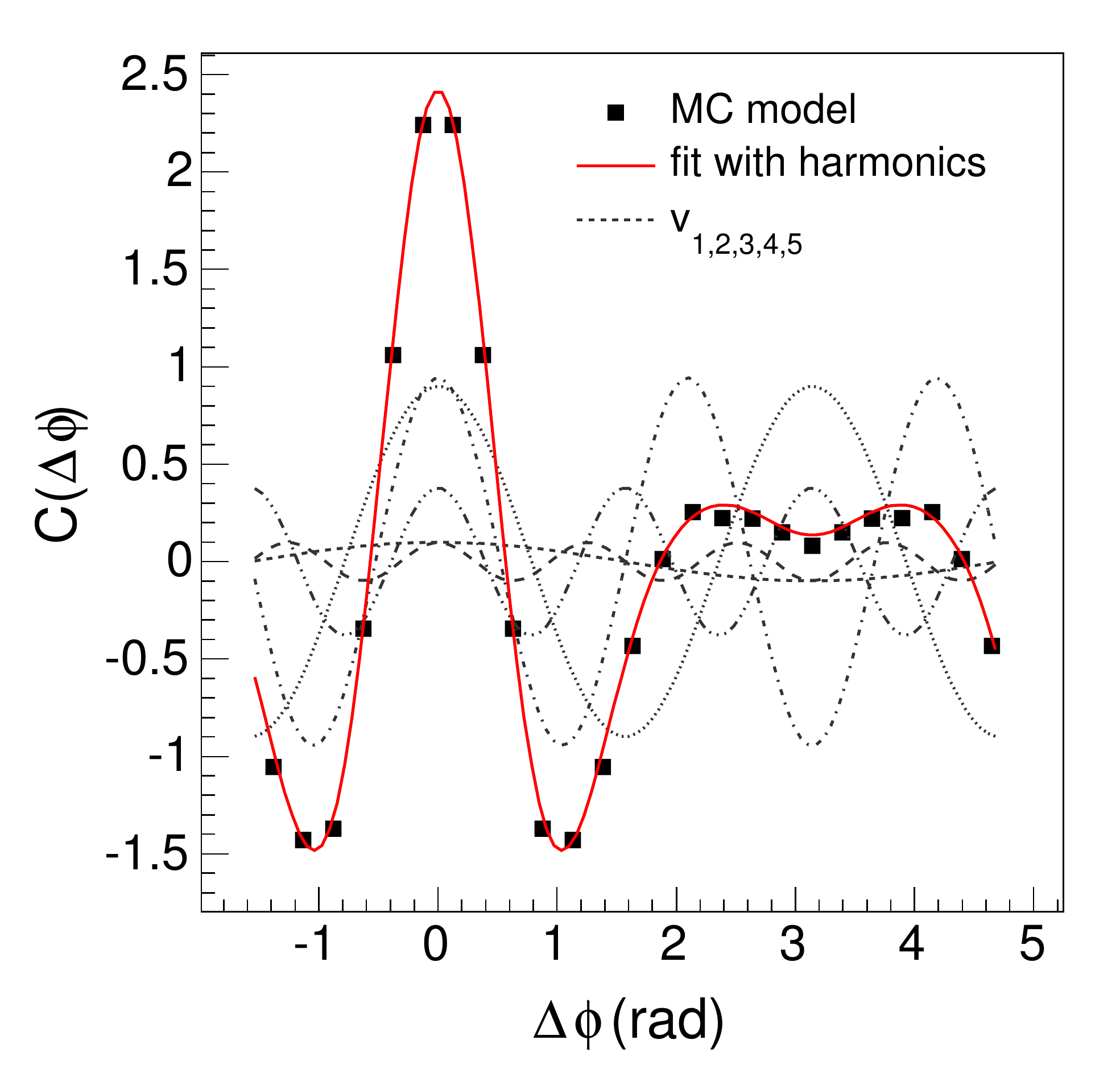}
	\put(60,65){ MC model }
	\put(54,22){ \small $\pt$ 3-5 GeV/{\it c} }
\end{overpic}
}
\hspace{-0.4cm}
\subfigure[a][]
{
\begin{overpic}[width=0.33\textwidth]{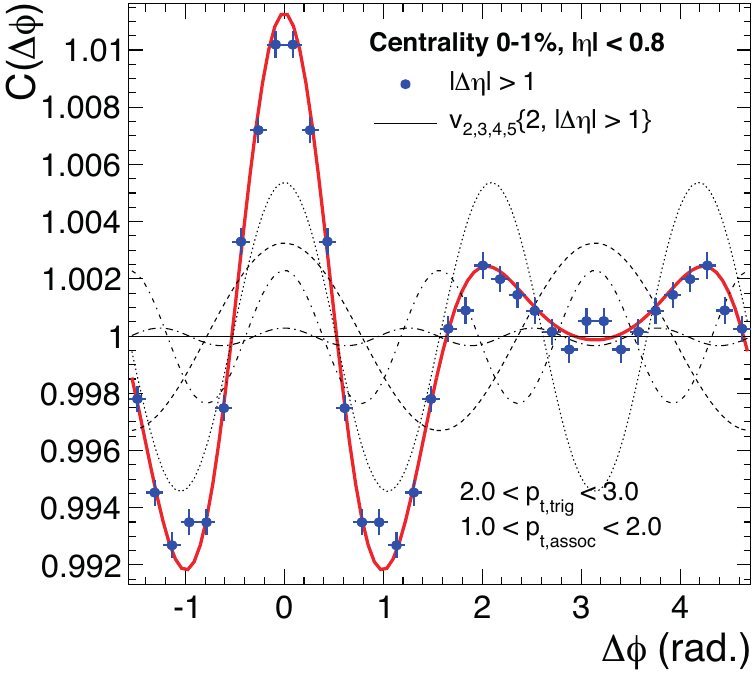}
	\put(74,65){ ALICE }
\end{overpic}
}
\caption{
(a) MC model results for  two-particle correlation function for very central nucleus-nucleus collision events for charged particles with $\pt$ in $[3, 5]$ GeV/$c$ \cite{IA,OK-IA-GF}. String interaction radius is 2 fm. (b)  Double-peak around $\Delta\vf=\pi$ in harmonics decomposition of two-particle angular correlations
in MC model at the most central Pb-Pb events  ($b=0$~fm) for charged particles with $\pt\in[3,5]$ GeV/{\it c}. The solid red line shows the sum of the  anisotropic flow Fourier coefficients $v_1$...$v_5$ (dashed lines).
(c) Experimental results \cite{ALICE-H}
for harmonics decomposition of two-particle angular correlations in  central Pb-Pb collisions at the LHC. Note: the different definitions for two-particle correlation function are used at (b) and (c).
}
\label{fig:v1234}
\end{figure}
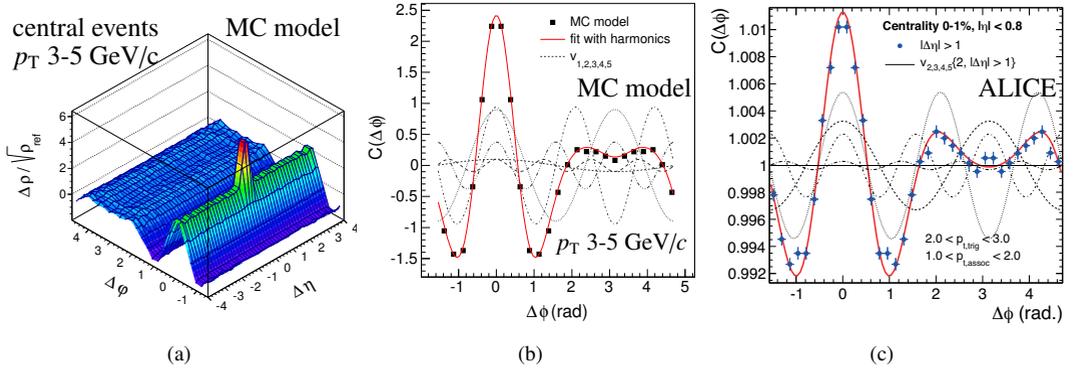

It was also obtained in \cite{OK-IA-GF} that 
 two-particle azimuthal correlations,
observed in very central Pb-Pb collisions at LHC \cite{ALICE-H},
can be obtained in the MC model with repulsive strings,
along with the
harmonic decomposition of the correlation function
(see Figure~\ref{fig:v1234}). This decomposition reveals 
significant values of $v_2$, $v_3$ and $v_4$ coefficients.

\section{Transverse momentum spectra and mass ordering }
\label{sec-3}

Another important experimental observation
is a so-called mass ordering of flow harmonics,
when $v_n(\pt)$ points  are "blue-shifted"
to the right for  hadrons with higher masses
(see, for example, results for elliptic flow in Pb-Pb collisions in \cite{alice_v2}).
It is natural that the  MC model should be able to describe 
spectra and flow coefficients for different kind of particles.
In the present work we  check a dependence $v_2$
on $\pt$ for pions, kaons and protons in order to see 
if the mass ordering is preserved.

\begin{figure}[H]
\centering
\sidecaption
\begin{overpic}[width=0.49\textwidth]{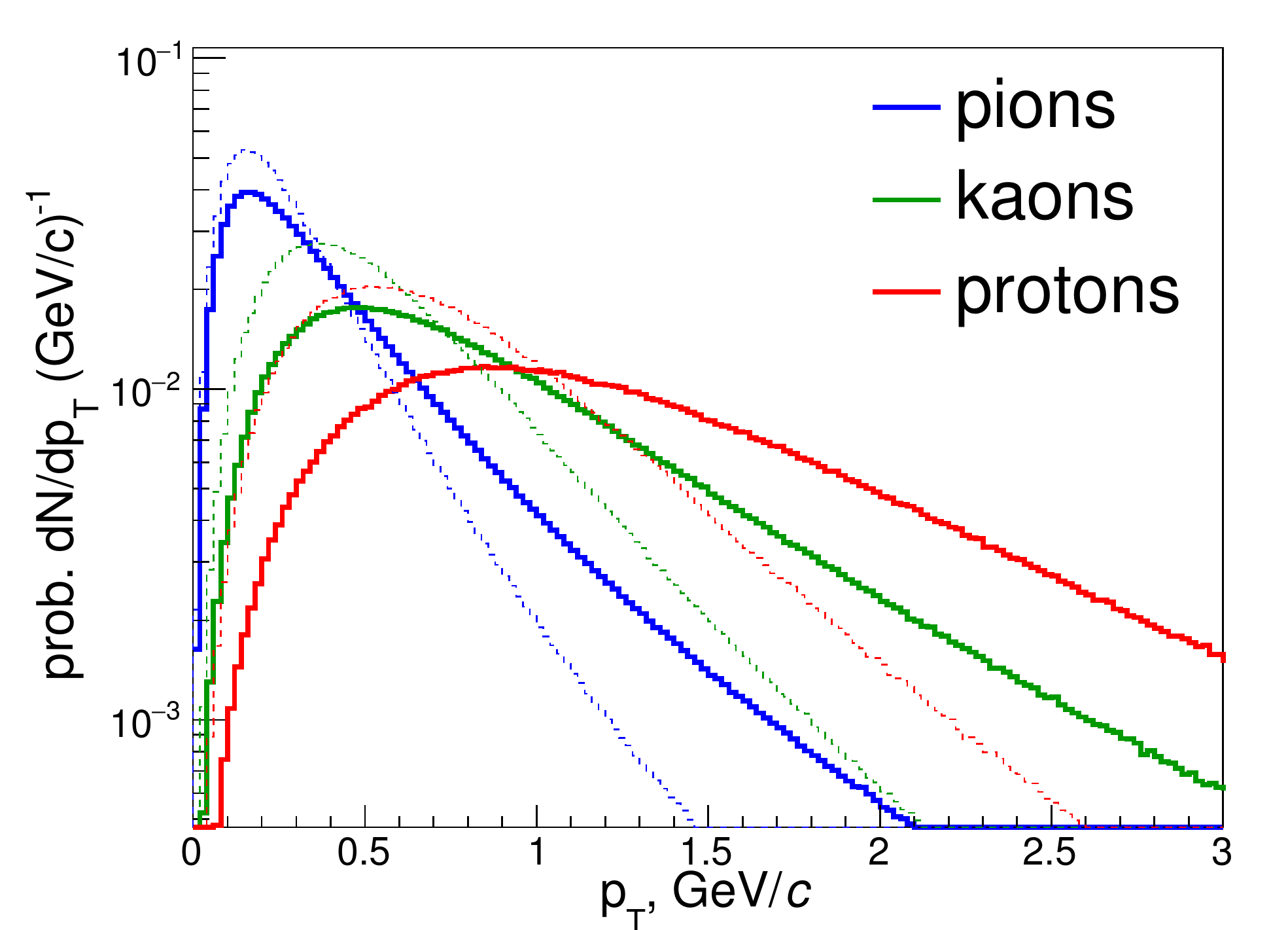} 
      \put(18,63){  MC model }
      \put(65,41){  dotted lines: }
      \put(65,36){  before boosts }\end{overpic}
\caption{
The $p_{\rm T}$-spectra for pions, kaons and protons 
in MC model  before  (dotted lines)
and after string boosts (solid lines),
in Pb-Pb collisions with centrality 10-20\%. 
}
\label{fig:spectra} 
\end{figure}

For this study, string fragmentation in  its rest frame was changed: 
pions, kaons and protons were generated as products of string decay (instead of just pions and $\rho$-mesons), and their
 transverse momentum spectra 
were parametrized by Tsallis formula tuned with data
for pp collisions at $\sqrt{s}=7$~TeV  \cite{alice_pp_pi_K_p}.
Obtained spectra 
are shown in Fig.\ref{fig:spectra} before application of string boosts (dotted lines),
and after boosting (solid lines).
Modifications of  spectra for different types of particles are found to be  in line with the observed hardening of $\av{\pt}$ observed for hadrons  for more central Pb-Pb events \cite{alice_PbPb_pi_K_p}.

\begin{figure}[H]
\centering
\subfigure[a][]
{
\begin{overpic}[width=0.49\textwidth,trim={0.2cm 0.9cm 1.4cm 0.2cm},clip]{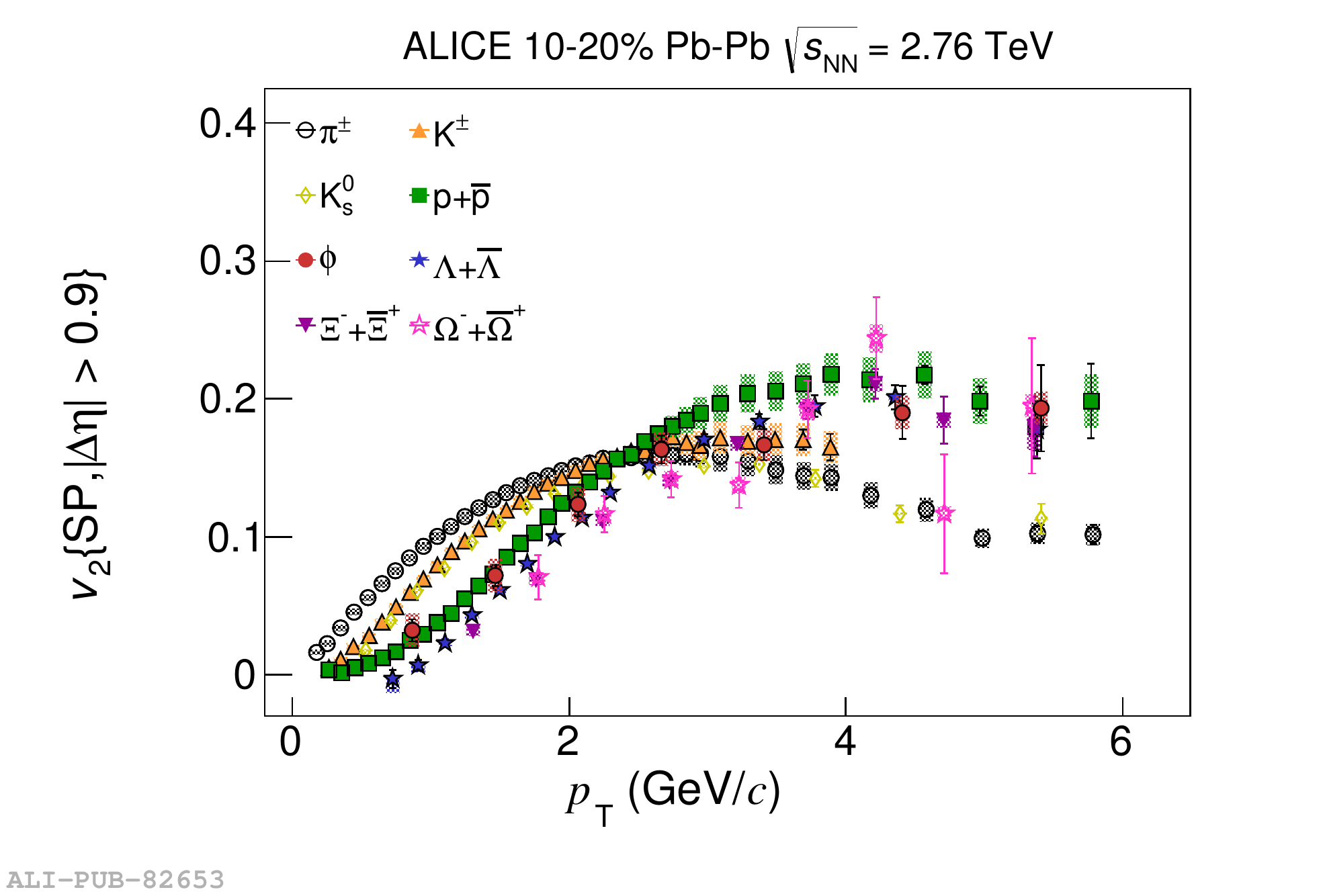} 
       \put(78,58){ \tiny ALI-PUB-82653 }
\end{overpic}
}
\hspace{-0.5cm}
\subfigure[a][]
{
\begin{overpic}[width=0.49\textwidth]
{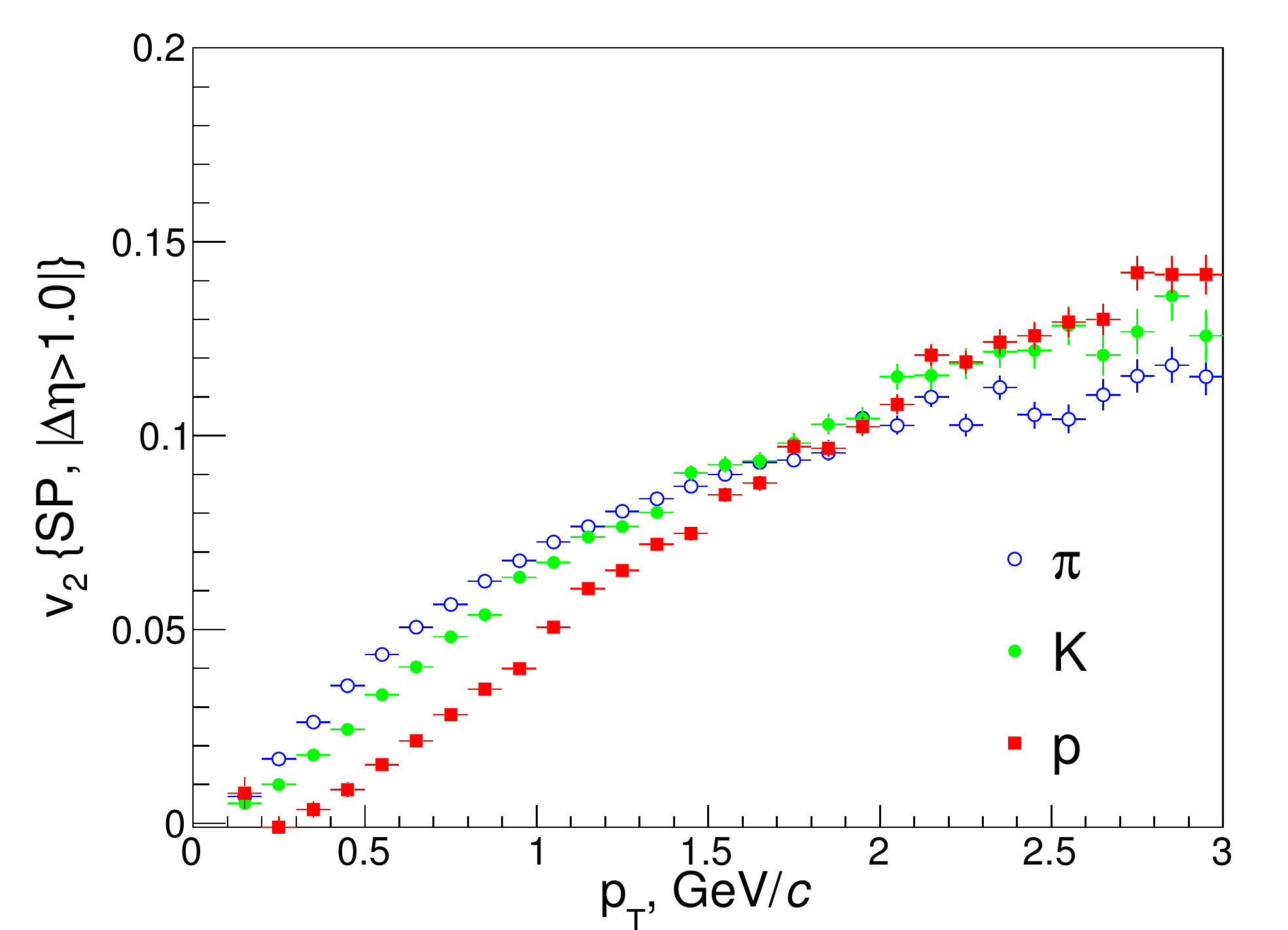} 
      \put(18,63){  MC model }
      \put(18,58){  10-20\% by impact par. }
\end{overpic}
}
\caption{
The $p_{\rm T}$-differential $v_2$ for pions, kaons and protons 
in centrality class 10-20\% of Pb-Pb collisions:\newline
(a) -- ALICE data at $\sNN=2.76$ TeV \cite{alice_v2},
(b) -- MC model calculations.
The string interaction radius $R_{int}=2$~fm.
}
\label{fig:mass_ordering} 
\end{figure}

To check the mass ordering,
the $v_2(\pt)$ was calculated  with the Scalar Product (SP) method, similarly  to what was done with LHC Pb+Pb data in \cite{alice_v2}. 
Hadrons of one of the species ($\pi$, K or p) in $|\eta|<1$  were chosen as particles-of-interest, while
reference charged particles were taken from $-3<\eta<-1$ and $1<\eta<3$, providing a pseudorapidity gap of $|\Delta\eta| > 1.0$.
This choice of rapidity ranges emulates experimental conditions 
of the $v_2(\pt)$  analysis of identified hadrons
 in ALICE experiment \cite{alice_v2}. 
The MC model results
are shown in Fig.\ref{fig:mass_ordering} (b) and compared to data (a).
It can be seen that the mass ordering observed in the experiment 
is qualitatively reproduced by the MC model in the region of interest (for $\pt$ values below 3 GeV/$c$).

Is should be mentioned, that 
$v_2(\pt)$ dependences for different particle species
are quite sensitive to the shape of the $\pt$ spectra,
and parameterization of the spectra by Tsallis distribution
tuned out to be good for reproduction of qualitative behaviour 
of the $v_2(\pt)$.
Another important point is 
that the effective string interaction radius $R_{int}$, which is the main parameter of the MC model,
is still needed to be quite large -- of about 2 fm.

\section{Discussion}
\label{sec-5}

Results of application of the MC model with the repulsive string-string interaction show that    quite a {\it  large effective string-string interaction radius} of about 2 fm is needed in order  to  describe qualitatively (and in some cases
almost quantitatively) azimuthal and rapidity topology of  two-particle correlations picture, including the harmonic decomposition, centrality dependence of flow coefficients, mass ordering, etc.  Possible physical mechanisms behind it  could be the following: \\
(i) existence of rather weak (but long-range in transverse plane) interactions between  color flux tubes (strings) 
mediated by exchange of some kind of meson, 
or \\
(ii) necessity to have  a "dynamical" phase  for a string when it moves
within "string medium", before decaying into hadrons. It  means that the string should  experience some pressure from other strings
(thus effectively increasing the interaction distances). 

The current version of the MC model is based on simple kinematic assumptions
and just a few  parameters.
There is  still an open question:  is it possible
to describe pp and p-Pb effects 
with minor changes in model assumptions and parameter values?
This check is be necessary to do if one wants
to establish some uniform mechanism of flow generation.

\section{Conclusions}
\label{sec-7}

High density string medium could be formed in hadronic collisions. 
It is shown in the Monte Carlo model that collective effects could  appear due to some, rather weak,  repulsive type of interaction between color flux tubes (strings).  The particle-emitting sources (strings) are boosted by the combined repulsion by neighbours, thus the initial space anisotropy is converted into 
the anisotropic azimuthal  distributions of charged particles.

Similar results could occur at sufficiently high string density  p-Pb and pp collisions.
More detailed quantitative analysis including mass ordering of flows  in p-Pb and pp collisions has to follow.

Acknowledgements. This work is supported by the Saint-Petersburg State University research grant 11.38.242.2015. The authors are grateful to V.Vechernin and V.Kovalenko for useful discussions.

\end{document}